\documentclass[journal=jpcafh]{achemso}
\setkeys{acs}{articletitle = true}

\usepackage{natbib}
\usepackage{lineno}
\usepackage{arydshln}
\usepackage{bibentry}
\usepackage{verbatim}
\usepackage{epsf}
\usepackage{color}
\usepackage{mathtools}
\usepackage{amssymb}
\usepackage{bm}
\usepackage{graphicx}
\usepackage{epstopdf}
\renewcommand{\eqref}[1]{(\ref{#1})}

\newcommand{\3}{$_{3}$}
\newcommand{\cm}{cm$^{-1}$}

\newcommand{\abinitio}{\textit{ab initio}}

\newcommand{\onlinecite}[1]{\citenum{#1}}
\newcommand{\p}{^\prime}
\newcommand{\pp}{^{\prime\prime}}

\newcommand{\ai}{\textit{ab initio}}
\mciteErrorOnUnknownfalse

\title[Line List for Methyl Radical]{ A Variationally Computed IR Line List for the Methyl Radical CH\3}

\author{Ahmad Y. Adam}
\affiliation[Bergische Universit{\"a}t]{Fakult{\"a}t f{\"u}r Mathematik und Naturwissenschaften, Physikalische
und Theoretische Chemie, Bergische Universit{\"a}t Wuppertal, D--42097 Wuppertal, Germany}
\author{Andrey Yachmenev}
\affiliation[DESY and Universit{\"a}t Hamburg]{Center for Free-Electron Laser Science, Deutsches Elektronen-Synchrotron DESY, Notkestra{\ss}e 85, D-22607 Hamburg, Germany,
The Hamburg Center for Ultrafast Imaging, Universit{\"a}t Hamburg, Luruper Chaussee 149, D-22761 Hamburg, Germany}
\author{Sergei N. Yurchenko}
\affiliation[University College London]{Department of Physics and Astronomy, University College London,
Gower Street, London WC1E 6BT, United Kingdom}
\author{Per Jensen}
\affiliation[Bergische Universit{\"a}t]{Fakult{\"a}t f{\"u}r Mathematik und Naturwissenschaften, Physikalische
und Theoretische Chemie, Bergische Universit{\"a}t Wuppertal, D--42097 Wuppertal, Germany}
\email{jensen@uni-wuppertal.de}

\date{\today}

\begin{document}

\begin{abstract}
We present the first variational calculation of a hot temperature \textit{ab initio} line list for
the CH$_3$  radical. It is based on a high level {\it ab initio} potential energy surface and dipole
moment surface of CH$_3$ in the ground electronic state. The ro-vibrational energy levels and Einstein
 $A$ coefficients were calculated using the general-molecule variational approach implemented in the
computer program TROVE. Vibrational energies and vibrational intensities are found to be in  very good
agreement with the available experimental data. The line list comprises 9,127,123 ro-vibrational states ($J\le 40$) and 2,058,655,166 transitions covering the wavenumber range up to 10000~cm$^{-1}$ and should be suitable for temperatures up to $T= 1500$~K.

\end{abstract}

\section{Introduction} \label{sec:intro}

The methyl radical CH$_3$ is a free radical of major importance in many areas of science such as hydrocarbon
combustion processes,\cite{comb90} atmospheric chemistry,\cite{atmo88}  the chemistry of
semiconductor production\cite{semi87}, the chemical vapor deposition of diamond\cite{diamond91},
and many chemical processes of current industrial and environmental interest. It is also expected to be present
in exo-planetary atmospheres \cite{astro02}, in the atmospheres of Saturn\cite{astro98} and Neptune\cite{astro99},
and in the interstellar medium,\cite{astro00} where it is thought to be one of the most abundant
free radicals.\cite{astro02} Because of its central role in this variety
of situations, its structural and spectroscopic parameters have been extensively studied.
Diverse spectroscopic techniques have been employed to determine  absolute concentrations of CH$_3$
in the gas phase, including UV/visible\cite{xelec03}, infrared\cite{nu2_72}, and Raman spectroscopies
\cite{nu1_84,nu1_88,nu1_92,nu1_94,nu1_96,dnu1_89a,dnu1_89b}.  In addition, CH$_3$ is an example of
a molecule with large vibrational contribution to the hyperfine coupling constant, accounting for up
to about 41\%\ of the total value (see Ref.~\citenum{aa15} and references therein).

Owing to the importance of CH$_3$ in various contexts, in particular in astrophysics and -chemistry,
its concentrations or column densities in remote environments such as interstellar space, the terrestrial
atmosphere, exo-planetary atmospheres, and the outer layers of cool stars are of interest and it is
desirable to determine these by remote-sensing spectroscopic methods. A prerequisite for such determinations
is the knowledge of the transition moments for the observed transitions, and these must often be obtained
in theoretical calculations as done, for example, in the ExoMol project\cite{jt528,exomol} by Yurchenko and
co-workers.\cite{jt564,nh3_09,jt571,jt592,jt580,jt641,jt597,jt612,jt701,jt620,jt638,jtlinTROVE,jt729}
This project aims at providing theoretically computed transition moments and simulated spectra for (small
to medium-sized) general polyatomic molecules of astrophysical and/or -chemical interest. In general, molecular transition moments known to within 10-20\%\ (the typical accuracy of the transition moment data of the HITRAN database\cite{GORDON20173}) are sufficiently accurate to be useful for most applications involving determination of concentrations and column densities, at least for  fundamental and overtone bands.  The level of \textit{ab initio} theory used in the present study [RCCSD(T)-F12b /cc-pVQZ-F12; see below] is expected to be sufficient to satisfy this requirement for the transition moments computed in the present work. Also, we expect the inaccuracy of our calculations to be predominantly caused by the inaccuracy of the \textit{ab initio} potential energy surface, and not by the truncation of the kinetic energy operator expansion.

At equilibrium, the three protons of electronic-ground-state $\tilde{X}$~$^2A''_2$
CH$_3$ form an equilateral triangle with the C nucleus at the centre-of-mass of the planar structure
with {\itshape\bfseries D}$_{3{\rm h}}$ point group symmetry (see Table~A-10 of Ref.\citenum{mss}).
There is no permanent dipole moment, and so the pure rotational transitions are dipole forbidden and
very weak. Also, the planar ground-state equilibrium structure precludes most one-photon transitions
to excited electronic states.\cite{xelec83} Owing to the extremely weak rotational spectrum, determinations
of concentrations and column densities for CH$_3$ must be made with rovibrational transitions in the
infrared region. The most suitable transitions are those in the intense $\nu_2$ fundamental band at
606~cm$^{-1}$ (where $\nu_2$ is the out-of-plane bending mode). This band provides convenient transitions
for concentration measurements and  has been used extensively for this purpose\cite{mu2_83,mu2_89,mu2_05,mu2_08}.
As mentioned above, the corresponding transition moments must be known in order that concentrations
can be determined, and the present work can be viewed as a first step towards providing extensive
catalogues of theoretical transition moments for CH$_3$, so-called line lists, of use in astrophysical
studies. In the present work we apply a high level \ai\ potential energy surface, refined by means
of experimental spectroscopic data, and \ai\ dipole moment surfaces to compute, with the TROVE program,
\cite{trove05,trove07,trove15,trove17a} sufficient energies and transition moments for generating
a hot ($T=1500$~K) IR line list for CH$_3$.

\section{Theory}

\subsection{Potential energy surface}\label{sec:PES}

The potential energy surface (PES) employed for the electronic ground state of CH\3\ in the present
work is based on the \ai\ surface reported in Ref.\citenum{aa15}, which we denote as PES-1. The PES-1
electronic energies were computed for 24\,000 symmetry-unique molecular geometries at
the open-shell RCCSD(T)-F12b~\cite{adler07,adler09} level of theory (explicitly correlated F12 restricted
coupled cluster included single and double excitations with a noniterative correction for triples) and the
F12-optimized correlation consistent polarized valence basis set cc-pVQZ-F12.\cite{peterson08}  The carbon inner-shell electron pair was treated as frozen core in the correlated
calculations. By using the frozen-core approximation we benefit from error cancellation. It is well known that, e.g. for second-row atoms, the core-valence correlation is almost exactly cancelled by the more costly high-order correlation effects.\cite{doi:10.1063/1.3624570}  Keeping only one of them would  make the accuracy deteriorate.

The analytical representation
for the PES was obtained in a least-squares fitting procedure using the parameterized function from {Lin {\it et al.}
\cite{nh3_02}}:

\begin{eqnarray}
V(\xi_1,\xi_2,\xi_3,\xi_{4a},\xi_{4b};{\sin\bar \rho}) \,
    &=&  V_{\rm e} +  V_0({\sin\bar \rho}) + \sum_j \, F_j({\sin\bar \rho}) \, \xi_j
\nonumber \\
& + &
       \sum_{j \leqslant  k} \, F_{jk}({\sin\bar \rho}) \, \xi_j \, \xi_k
       +  \sum_{j \leqslant k \leqslant l} \, F_{jkl}({\sin\bar \rho}) \,\xi_j \, \xi_k \, \xi_l
\nonumber \\
       &+ & \sum_{j \leqslant k \leqslant l \leqslant m} \, F_{jklm}({\sin\bar \rho}) \, \xi_j \, \xi_k \,
        \xi_l \, \xi_m + \dots;
\label{e:PEF:morbid}
\end{eqnarray}
this function depends on the stretching variables
\begin{equation}
\xi_k =
1 - \exp\left[ -a ( r_k - r_{\rm e}) \right], \;\; k=1,2,3,
\end{equation}
where
 $r_k$ is the instantaneous value of the distance
between the C nucleus
 and the proton H$_k$
labeled $k$ $=$ 1, 2, or 3;
$r_{\rm e}$ is the common equilibrium value of
the three $r_k$ bond lengths,
and $a$ is a Morse parameter.
Furthermore,
the symmetrized bending variables
$( \xi_{4a}, \xi_{4b} )$ are defined as
\begin{equation}
\label{S4a4b}
\left( \xi_{4a}, \xi_{4b} \right) =
\left(
  \frac{1}{\sqrt{6}} [2 \, \alpha_{1} - \alpha_{2} - \alpha_{3} ],
 \frac{1}{\sqrt{2}} [\alpha_{2} - \alpha_{3} ]
\right)
\end{equation}
with
$\alpha_i$ as      the bond angle $\angle$(H$_j$XH$_k$) where $(i,j,k)$ is a
permutation of the numbers (1,2,3).
Finally,
the variable
\begin{equation}
\label{e:sin:brho} \sin {\bar \rho} \, = \, \frac{2}{\sqrt{3}} \,
\sin [
   (\alpha_1 + \alpha_2 + \alpha_3) /6] \hbox{}
\end{equation}
describes  the out-of-plane bending.
At the planar equilibrium configuration, we have
   $\alpha_1 + \alpha_2 + \alpha_3$ $=$ 360$^\circ$ and so
$\sin {\bar \rho}$ $=$
$\sin {\bar \rho}_{\rm e}$ $=$ 1.
The functions $V_0({\sin\bar \rho})$ and  $F_{jk\dots}({\sin\bar \rho})$
in Eq.~\eqref{e:PEF:morbid} are defined as
\begin{eqnarray}
\label{V0}
   V_0({\sin\bar \rho}) &=& \sum_{s=1}^{4} \, f_0^{(s)} \, (1           - \sin {\bar \rho})^s \hbox{,} \\
\label{Ffunction}
  F_{jk\dots}({\sin\bar \rho}) &=& \sum_{s=0}^{N}\, f_{jk\dots}^{(s)} \, (1 - \sin {\bar \rho})^s \hbox{,}
\end{eqnarray}
where
the quantities  $f_0^{(s)}$ and $f_{jk\dots}^{(s)}$ in
Eqs.~(\ref{V0}) and~(\ref{Ffunction}) are expansion coefficients.
 The optimized values of the parameters
$a$,
$r_{\rm e}$,
 $f_0^{(s)}$, and $f_{jk\dots}^{(s)}$
are given in the supplementary material to
 Ref.~\citenum{aa15} together with
 Fortran~90 routine for calculating PES values.

The analytical form of PES-1\cite{aa15}
is given in terms of
 the \abinitio\ cc-pVQZ-F12 values of the
equilibrium structural parameters,
$r_{\rm e} = 1.0774$~\AA\ and $\alpha_{\rm e}=
120^\circ$,\cite{aa15}
for the electronic ground state of CH\3. In the present work, we optimized
the value of
 $r_{\rm e}$ in a least-squares fitting to experimentally derived
to rotational energy spacings within
 the vibrational states of CH\3.
The fitting produced
$r_{\rm e} = 1.0763$~\AA;
$\alpha_{\rm e}$ $=$ 120$^\circ $ by symmetry.
We use these optimized values of the equilibrium structural parameters.
All results presented below are based on
the analytical potential energy function called PES-2, obtained from PES-1\cite{aa15}
by replacing the
 \abinitio\ cc-pVQZ-F12 value of
$r_{\rm e}$  by
the adjusted value given here. The remaining PES-2 parameter values are identical to
those of PES-1 and can be obtained from
the supplementary material to Ref.~\citenum{aa15}.

\subsection{Dipole moment surface}\label{sec:DMS}

The dipole moment surface (DMS) for the electronic ground state of CH$_3$ was computed using the
MOLPRO\cite{molpro12}  program package. Frozen-core calculations were carried out for 19\,361
symmetry-unique geometries (15\,600 below 30\,000~\cm) using the spin-restricted open-shell coupled
cluster theory RCCSD(T)~\cite{watts} and the augmented correlation consistent valence basis set
aug-cc-pVTZ~\cite{dunning,kendall}, employing the two-point stencil central
finite differences with the electric field strength of 0.002~a.u.

We employ the so-called symmetrized molecular-bond (SMB) representation\cite{nh3_09}
[which is an extension of the
 molecular-bond (MB)
representation~\cite{nh3_05}]
to formulate analytical functions describing the molecular dipole moment
components.
The SMB representation makes use of
the projections
 $ \bar{\bm{\mu}} \cdot \mathbf{e}_k $
of the dipole moment on the molecular
bonds, where
  $\mathbf{e}_k$ is  the unit vector along the
C--H$_k$ bond,
\begin{equation}\label{e:e:r}
   \mathbf{e}_k = \frac{\mathbf{r}_k-\mathbf{r}_4}{|\mathbf{r}_k-\mathbf{r}_4|}
\end{equation}
with $\mathbf{r}_k$, $k$ $=$ 1, 2, 3, as the position vector of proton $k$ and
$\mathbf{r}_4$ as the position vector of the C nucleus.

We form                                           symmetry-adapted
linear combinations of the MB projections $ \bar{\bm{\mu}} \cdot
\mathbf{e}_j $:
\begin{eqnarray}
\label{e:mu:A}
  \bar\mu_{A_2\pp}^{\rm SMB}&=& \left( \bar{\bm{\mu}} \cdot \mathbf{e}_{\rm N} \right) \\
\label{e:mu:Ea}
  \bar\mu_{E_a\p}^{\rm SMB} &=& \frac{1}{\sqrt{6}} \left[ 2 \left( \bar{\bm{\mu}} \cdot \mathbf{e}_1 \right) - \left( \bar{\bm{\mu}} \cdot \mathbf{e}_2 \right) - \left( \bar{\bm{\mu}} \cdot \mathbf{e}_3 \right) \right] \\
\label{e:mu:Eb}
  \bar\mu_{E_b\p}^{\rm SMB} &=& \frac{1}{\sqrt{2}} \left[                                                \left( \bar{\bm{\mu}} \cdot \mathbf{e}_2 \right) - \left( \bar{\bm{\mu}} \cdot \mathbf{e}_3 \right) \right],
\end{eqnarray}
where, in addition to the vectors
   $\mathbf{e}_k$,
 we have introduced
$\mathbf{e}_{\rm N}^{}$ $=$ $\mathbf{q}_{\rm N}^{}/\vert
\mathbf{q}_{\rm N}^{}\vert$ with
$\mathbf{q}_{\rm N}$ as
 the `trisector'
\begin{equation}
\label{e:q-N}
    \mathbf{q}_{\rm N} =
      (\mathbf{e}_1 \times \mathbf{e}_2)
    + (\mathbf{e}_2 \times \mathbf{e}_3)
    + (\mathbf{e}_3 \times \mathbf{e}_1).
\end{equation}
The
subscripts
$\Gamma$ $=$
$A_2\pp$, $E_a\p$, and $E_b\p$
of the quantities $\bar\mu_{\Gamma}^{\rm SMB}$
in Eqs.~(\ref{e:mu:A})--(\ref{e:mu:Eb}) refer to
the irreducible representations
 (Table~A-10 of Ref.~\citenum{mss})
of the CH$_3$ molecular symmetry group
{\itshape\bfseries D}$_{3{\rm h}}$(M);
the electronically averaged dipole moment
$ \bar{\bm{\mu}}$ generates the representation
$A_2 \pp$ $\oplus$ $E'$.
 The quantity $\bar\mu_{A_2\pp}^{\rm SMB}$ is
antisymmetric under the inversion operation\cite{mss} $E^*$ and vanishes
at planarity, so that
$ \bar{\bm{\mu}}$ has only two non-vanishing, linearly independent components
at planarity. These two components vanish at planar configurations with
{\itshape\bfseries D}$_{3{\rm h}}$ point group symmetry.

\begin{figure}[h]
\begin{center}
\caption{\label{fig:rms-DMS} Root-mean-square (RMS) errors of the fittings to the \ai\ dipole moment values.
The results of a series of fittings are shown. In each fitting, the data set includes the \ai\ points with
electronic energy (relative to the potential energy minimum) $V$ $\leqslant$ $E$ (see text).}
\includegraphics[width=0.9\textwidth]{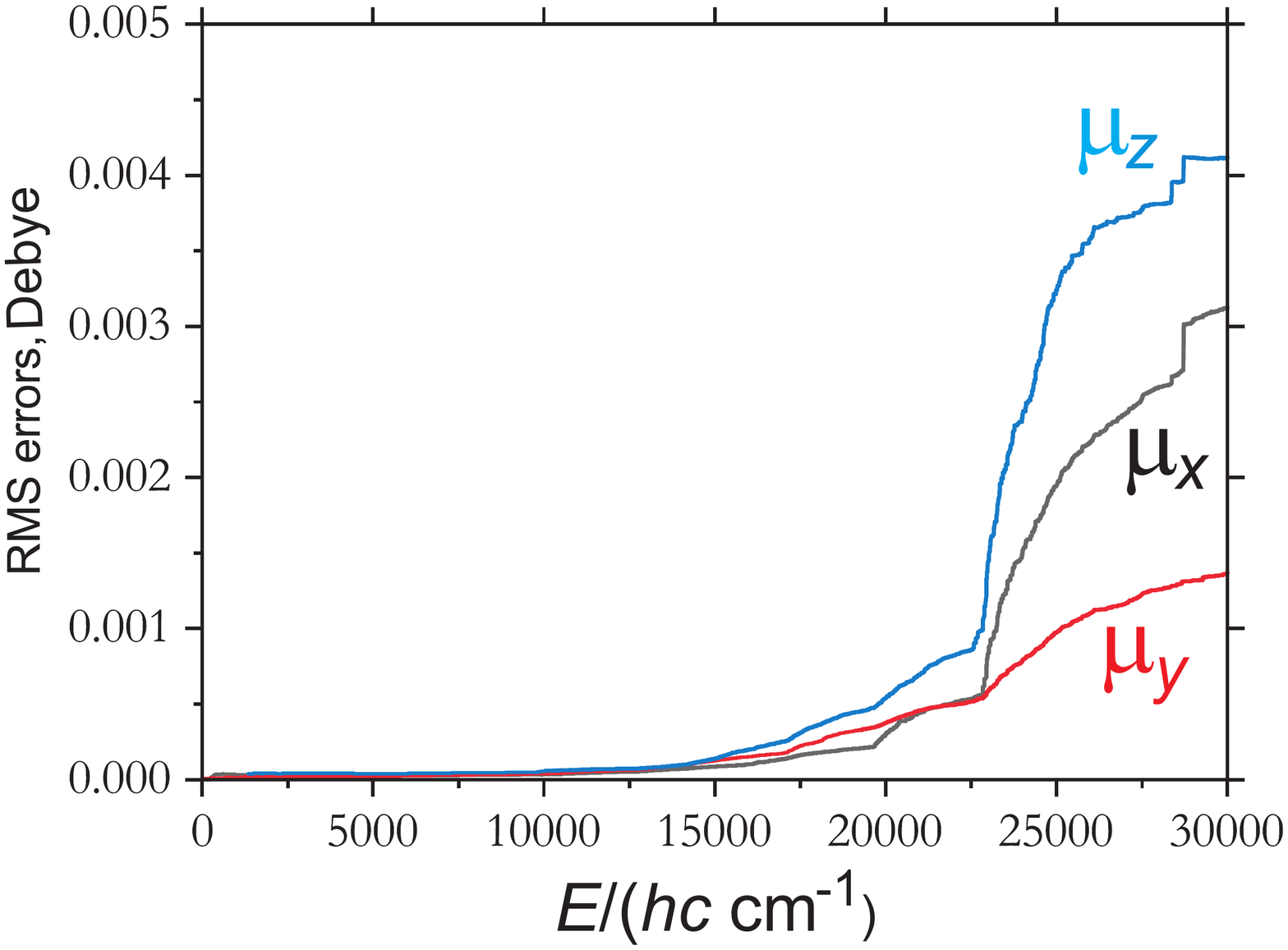}
\end{center}
\end{figure}

The three components of the SMB dipole moment
in Eqs.~(\ref{e:mu:A})--(\ref{e:mu:Eb}) are
 represented by 4$^{\rm th}$ order polynomial expansions
\begin{equation}\label{eq:dms_expansion}
\begin{aligned}
\bar{\mu}_\Gamma^{\rm SMB}
(\chi_1,\chi_2,\chi_3,\chi_{4a},\chi_{4b};\rho)= & \mu_0^{\Gamma}(\sin\bar\rho)+\sum_{i}\mu_i^{\Gamma}(\sin\bar\rho)\chi_i + \sum_{i\le j}\mu_{ij}^{\Gamma}(\sin\bar\rho)\chi_i\chi_j\\ + \sum_{i\le j\le k}\mu_{ijk}^{\Gamma}(\sin\bar\rho)\chi_i\chi_j\chi_k
 & + \sum_{i\le j\le k\le l}\mu_{ijkl}^{\Gamma}(\sin\bar\rho)\chi_i\chi_j\chi_k\chi_l ,
\end{aligned}
\end{equation}
in terms of the variables
\begin{equation}
\label{eq:mu:icoord1}
\chi_k = \Delta r_k (1-\exp(-\Delta r_k))^2, \quad (k=1,2,3)
\end{equation}
with $\Delta r_k$ $=$ $r_k$ $-$ $r_{\rm e}$
and
\begin{equation}
\label{eq:mu:icoord2}
( \chi_{4}, \chi_{5} ) =
( \xi_{4a}, \xi_{4b} )
\end{equation} where
$( \xi_{4a}, \xi_{4b} ) $ are defined in Eq.~\eqref{S4a4b}.
The expansion coefficients $\mu^\Gamma_{ij...} (\sin\bar\rho)$ are defined as
\begin{equation}
 \mu^\Gamma_{ij...}(\sin\bar\rho)=\sum_{s=0}^N\mu^{\Gamma(s)}_{ij...}(1-\sin\bar\rho)^s,
\end{equation}
where $\sin {\bar \rho} $ is given by Eq.~\eqref{e:sin:brho} and
the maximal order of the polynomial is $N=8$.
For more details the reader is referred to  Ref.\citenum{nh3_09}.
The final fit of the 15\,600  geometries required a total number of 218  parameters
(131 for
  $\bar\mu_{A_2\pp}^{\rm SMB}$
 and 87 for
  $\bar\mu_{E_a'}^{\rm SMB}$ and
  $\bar\mu_{E_b'}^{\rm SMB}$)
 and reproduced the \textit{ab initio} data with a root-mean-square (RMS) differences of 0.003~D, 0.001~D and 0.004~D for the $x,y$ and $z$ components respectively and  energies up to 30\,000 cm$^{-1}$, see Figure~\ref{fig:rms-DMS}. A series of fittings to the $x$, $y$ and $z$ \ai\ dipole-moment components have been carried out, including in the data set for each fitting the \ai\ points with
electronic energy (relative to the potential energy minimum) $V$ $\leqslant$ $E$, and increasing $E$. Figure~\ref{fig:rms-DMS} shows the RMS deviation for each dipole-moment component as a function of $E$.
The DMS expansion parameter set and the Fortran 90 functions are included in the supplementary material.

\subsection{Intensity simulations with TROVE} \label{sec:simulat:range}


The `readiness' of the molecule to make an absorption or emission transition from an initial ro-vibrational state $i$ to a final ro-vibrational state $f$ is expressed by the  line strength\cite{dipole05,mss,fms}  $S(f\leftarrow i)$, a  quantity with units of [dipole moment]$^2$ (typically Debye$^2$). For $S(f\leftarrow i)$ $=$ 0, the transition does not take place and need not be considered. As discussed in Ref.\citenum{mss} transitions with $S(f\leftarrow i)$ $\ne$ 0 are said to satisfy selection rules which we can derive from symmetry considerations before we do quantitative, numerical calculations of $S(f\leftarrow i)$. Thus, these calculations need only be done for transitions satisfying the selection rules, and after obtaining values of $S(f\leftarrow i)$ for these transitions, we can compute the corresponding  Einstein coefficients
and absorption intensities.

The initial(final) state $i$($f$) has the rotation-vibration wavefunction $\vert \Phi_{\rm rv}^{(i)}
\rangle$($\vert \Phi_{\rm rv}^{(f)} \rangle$).
The line strength\cite{dipole05,mss,fms} $S(f\leftarrow i)$ of the ro-vibrational transition
$f \leftarrow i$ is
\begin{equation}
\label{e:linestrength_deg}
S(f \leftarrow i) =
g_{\rm ns} \,
\sum_{M_f, M_i} \,
\sum_{A=X,Y,Z}
\left\vert  \left\langle
\Phi_{\rm rv}^{(f)} \,
\left\vert
\bar\mu_A \right\vert
\Phi_{\rm rv}^{(i)}
\right\rangle \right\vert^2 \hbox{,}
\end{equation}
where  the nuclear spin statistical weight factor\cite{mss} is denoted  $g_{\rm ns}$ and the electronically averaged component of the molecular dipole moment along the space-fixed axis\cite{mss} is denoted $\bar\mu_A$,
$A$ $=$ $X$, $Y$, or $Z$. The quantity $M_i$($M_f$) is the quantum number
defining the projection of the total angular
momentum $\hat{\mathbf{J}}$ on the $Z$ axis for the initial(final) state.

Assuming that the molecules considered are in thermal equilibrium at the absolute temperature $T$,
the intensity of a spectral line  is determined as
\begin{equation}
\label{e:intensity-absorption}
I(f \leftarrow i)  =
    \frac{8 \pi^3 N_A \tilde\nu_{if}}{(4 \pi \epsilon_0)3hc} \,
    \frac{{\rm e}^{-E_i/kT}}{Q} \,
    \big[1 - {\rm exp}(-hc\tilde\nu_{if}/kT)\big] \,
    S(f \leftarrow i).
\end{equation}
Here, the absorption wavenumber is denoted $\tilde{\nu}$, and Eq.~\eqref{e:intensity-absorption} yields the intensity of a
transition from the initial state $i$ with energy $E_i$ to the final state $f$ with energy $E_f$, where
$hc\tilde\nu_{if}$ = $E_f$$-$$E_i$.  The partition function $Q$ is  defined as $Q$ $=$ $\sum_j g_j \,
\exp (  -E_j/kT)$, where $g_j$ is the total degeneracy of the state with energy $E_j$ and
the sum runs over all energy levels of the molecule, and other symbols
have their usual meanings. The total degeneracy $g_j$ is given by
$(2J+1)$ times the spin degeneracy ($2S+1 = $ 2) and times the nuclear spin degeneracy which is 4, 0, 2, 4, 0, 2
for $A_1\p, A_2\p, E\p, A_1\pp, A_2\pp,$ and $E\pp$ symmetries respectively. The ground electronic
state of CH$_3$ is a doublet  ($\tilde{X}$~$^2A''_2$) with a small splitting\cite{nu2_82c,nu3_97b}
in the rovibrational energy levels due to spin-rotation interactions, around 0.01 \cm, which we
therefore chose to ignore in the present work.


Yurchenko~\textit{et al.}~\cite{dipole05} have given, in their Eq. (21),
a detailed expression for the
line strength of an individual
ro-vibrational transition within an isolated electronic state
of an XY$_3$ pyramidal molecule.
Assuming that the populations of the lower (initial) states are Boltzmann-distributed, we limit the intentity calculations to transitions starting from
levels below $E_{i}^{\rm max}/h c = $ 9\,000~\cm. With this limitation,
 Boltzmann factors of
$\exp(-E_i / k T) > 2\times 10^{-4}$ enter into Eq.~\eqref{e:intensity-absorption} for $T=1500$~K. It is common to use the partition function for estimating the completeness of the line list for a given temperature \cite{jt592}. Towards this end, we consider the ratio $Q_{9000\,{\rm cm}^{-1}}/Q_{\rm total}$, where $Q_{\rm total}$ is the converged partition function value calculated by explicit summation over all computed  energy levels and $Q_{9000\,{\rm cm}^{-1}}$ is the partition function value calculated by summation over levels with energies lower than 9000~\cm. This ratio gives 95~\%\ completeness
at temperatures below 1500~K. Consequently, we estimate $T$ $=$ 1500~K to be the maximal temperature for which our line list is realistic. Since it is safe to limit the lower-state energies to be below 9000~\cm, it is sufficient to consider rotational states with $J$ $\leqslant $40. We compute a line list in the wavenumber range  0--10\,000~\cm; the upper energy limit (i.e., the maximum value of
the final-state energy) corresponds to a term value of $E^{\rm max}/hc $ $=$ 19,000~\cm.

\section{ Computational details} \label{subs:Computational-details}
The variational nuclear-motion
calculations are done with a symmetry-adapted basis set. With such a basis set, the Hamiltonian matrix becomes block diagonal according to the irreducible representations of the {\itshape\bfseries D}$_{3{\rm h}}$(M) molecular symmetry group:\cite{mss} $A_1\p$,
$A_2\p$, $A_1\pp$, $A_2\pp$, $E\p$, and $E\pp $. The
$A_2\p$ and $A_2\pp$ matrices are of no interest for CH\3 as the
corresponding states have zero nuclear spin statistical weights and do not exist in nature.\cite{mss} 
The $E\p$ and $E\pp $ matrices each split into two sub-blocks, of which only one must be diagonalized.\cite{mss}

The calculation of the matrix elements
$\langle \Phi_{\rm rv}^{(f)} \, \vert \bar\mu_A \vert \Phi_{\rm
rv}^{(i)} \rangle$ in Eq.~\eqref{e:linestrength_deg} is the
bottle-neck in the spectrum simulations. Here, the wavefunctions $\Phi_{\rm rv}^{(w)} $
are given as superpositions of symmetry-adapted  basis functions (see
Eq.~(65) of Yurchenko \textit{et al.}~\cite{trove17a}):
\begin{equation}
\label{e:rovib:basis}
\vert \Phi_{\rm rv}^{(w)} \rangle =
\sum_{V K \tau_{\rm rot}}
C_{V K \tau_{\rm rot}}^{(w)} \,
\vert J_w \, K \, m_w\, \tau_{\rm rot}  \rangle \, | V \rangle,
\;\; w = i \;\; {\rm or} \;\; f,
\end{equation}
with the $C_{V K \tau_{\rm rot}}^{(w)}$ as expansion coefficients.
In Eq.~\eqref{e:rovib:basis}, the symmetrized rotational basis functions are
denoted
$\vert J_w \, K \, m_w\, \tau_{\rm rot} \rangle$ with $\tau_{\rm rot}$ ($=$ 0
or 1) defining the rotational parity, and $\vert V \rangle$ is a vibrational basis function.
In order to accelerate this part of the calculation, we pre-screened the expansion coefficients $C_{V
K \tau_{\rm rot}}^{(f)}$.
All
terms with coefficients less than the threshold value of
$10^{-13}$ were discarded in the intensity calculation.

The evaluation of the dipole moment
matrix elements $\langle \Phi_{\rm rv}^{(f)} \, \vert \bar\mu_A
\vert \Phi_{\rm rv}^{(i)} \rangle$ has been made more efficient in a
two-step procedure. In the first step, an effective line strength is evaluated for a given lower state $i$:
\begin{equation}\label{e:effective:Sif}
S_{i,V K}^{A} = \langle \Phi_{\rm rv}^{(i)} \vert \bar\mu_{A} \vert \phi_{V K} \rangle.
\end{equation}
Here,   $\phi_{V K}$ is a short-hand notation for the
primitive basis function $\vert J_w K \, m_w\, \tau_{\rm rot}
\rangle$ $\times \vert V \rangle $. From the $S_{i,V K}$-values obtained,
we compute,
in the second step, the line strength $S(f \leftarrow i)$  as
\begin{equation}\label{e:effective:Sif:2nd:step}
S(f \leftarrow i)= g_{\rm ns} \sum_{m_i, m_f}\sum_{A=X,Y,Z} \left\vert  \sum_{V, K} C_{V K \tau_{\rm rot}}^{(f)} S_{i, V K}^A \right\vert^2.
\end{equation}

We had to compute a very large number of transitions satisfying the selection rule
$\vert J_{f} - J_i \vert$ $\leqslant$ 1, where
 $J_{i}$ and $J_f$ are the values of the angular momentum quantum number  $J$ for the initial and final state, respectively. Consequently, we saved memory by organizing the calculation of the ro-vibrational eigenstates and the $S(f\leftarrow i)$-values such that at a given time, only eigenvectors for states with two consecutive $J$-values,  $J$ and $J+1$, are available for the computation of $S(f\leftarrow i)$-values. This algorithm is  implemented in the  GPU GAIN-MPI program.\citep{17AlYuTe}

The vibrational basis set $|V\rangle$ is obtained in TROVE using a multi-step contraction and symmetrization procedure, starting from local primitive basis set functions, each depending on one variable only (see Refs.~\citenum{trove07,trove15,trove17a} and references therein).
Thus, a compact representation of the vibrational basis set is obtained in a form
 optimized for the molecule of interest.
 The final vibrational basis set is represented by the eigenfunctions of the purely vibrational part of the Hamiltonian;
we call these eigenfunctions the `$J=0$ basis'.

\section{Results} \label{sec:trove}

\subsection{Basis set convergence and empirical adjustment of the vibrational band centers}
\label{subs:band-centers:adjustment}
The dimensions of the Hamiltonian matrix blocks to be diagonalized
are important in determining
the accuracy of the computed
energies and wavefunctions for highly excited ro-vibrational states.
  Consequently it is imperative to determine empirically the smallest basis set with which the
 required eigenvalue accuracy (i.e., the
optimum basis-set size for `convergence') can be attained.

In TROVE, the size
of the vibrational basis set is controlled by polyad number
truncation.\cite{trove05,trove07,trove15} For CH$_3$, the polyad number $P$ is defined as:
\begin{equation}\label{e:polyad-2}
P =  2 (n_1 + n_2 +n_3) + n_{4} + n_{5} + n_{6},
\end{equation}
where $n_i$ are the principal quantum numbers associated with the primitive
functions $\phi_{n_i}(\xi_i)$. The primitive vibrational basis functions are products of one-dimensional basis functions $\phi_{n_i}(\xi_i)$, and only products with $P$ $\leq$ $P_{\rm max}$ are included in the primitive vibrational basis.

An even tighter level of convergence could be achieved for the vibrational term values if these were calculated with different  $P_{\rm max}$-values and the resulting progression of term values were extrapolated to the complete vibrational basis set limit~\cite{ph3_08}.
However, for the purpose of generating
line lists this is not considered necessary. The corrections from the extrapolation will be
small compared with the term-value errors caused  by the imperfection of the underlying potential energy surface.
Instead, we pragmatically aim for a higher accuracy by
 resorting to an  empirical approach: The theoretical vibrational
term values are replaced by the available accurate,
 experimentally derived  vibrational band-centre values.
In this manner, we are adjusting the vibrational band centers `manually'; this empirical adjustment also shifts the rotational energy-level structure towards better
agreement with experiment.
We call this procedure the EBSC scheme as it can be regarded as an Empirical Basis Set Correction.

 We adopt the EBSC scheme for the vibrational bands $\nu_2$, $2\nu_2$, $\nu_1$, $\nu_4^1$, and $\nu_3^1$, for which accurate experimental data are available, in combination with PES-2, where we have adjusted the equilibrium structure of the molecule to fit the experimentally derived pure rotational term values. The vibrational basis set was truncated at the polyad number $P_{\rm max}=32$.  We incorporate experimental information in the
EBSC scheme, and so we obviously depart from a purely \abinitio\ approach. This
 is considered justified by the accuracy improvement that can
be achieved in the computation of an extensive ro-vibrational line list.

To improve the accuracy of the predicted
vibrational band-centers, a more thorough refinement of the
PES would be required. However, the available accurate experimental data for  the vibrationally excited states of CH$_3$ is severely limited, and so we opted for the EBSC approach in conjunction with the $r_{\rm e}$-refinement.
For all bands that are not EBSC-corrected, the predicted vibrational term values are determined to a significant extent by the \textit{ab initio} data, and so their accuracy is limited. However, we have improved the prediction of the rotational structures, and that will facilitate the assignments of future experimental spectra for CH$_3$.

 In Table~\ref{t:energies}, the vibrational term values below 5000~\cm\ of the methyl radical, calculated variationally in the present work from PES-2, are compared with the available experimental data.
 The EBSC substitution was made in the
$J>0$ TROVE calculations of the present work, in that
the theoretical vibrational term values (obtained for
 $P_{\rm max} = 32$) were replaced by the experimental values in Table~\ref{t:energies}. This table also shows the effect of the polyad number $P_{\rm max}$ on the
vibrational energy. 



\begin{table}
\caption{\label{t:energies}  Vibrational band centers (\cm) of $^{12}$CH\3\ from variational calculations.}
\begin{center}
\tabcolsep=5pt
\renewcommand{\arraystretch}{1.175}
\begin{tabular}{llcr@{}lrrr@{}l}
\hline
\hline
$\Gamma$& State &  Ref. & \multicolumn{2}{c}{Obs.$^a$}   & \multicolumn{1}{c}{$P_{\rm max}=24$$^b$}  & \multicolumn{2}{c}{$P_{\rm max}=32$$^c$} &\\
\hline
$A_1'$  &       $       2\nu_2                  $       & \onlinecite{nu2_81} 	& 1288.1\phantom{00}    &       &       1279.77	&	1281.24	&	\\
        &       $       2\nu_4                  $	& 	                &		        &	&	2737.63	&	2739.64 &	\\
        &	$	4\nu_2	                $	& 	                &	 	        &	&	2773.65	&	2776.86	&	\\
        &	$	 \nu_1	                $	& \onlinecite{nu1_92}   & 3004.42\phantom{0}    &       &       3002.71 &       3002.76 &       \\
        &	$	3\nu_4^3	        $	& 	                &		        &	&	4118.59	&	4120.58	&	\\
        &	$	 \nu_1+2\nu_2	        $	&	                &		        &	&	4258.97	&	4260.53	&	\\
        &	$	6\nu_2	                $	&	                &		        &	&	4391.99	&	4397.00	&	\\
        &	$	 \nu_3^1+\nu_4^1	$	&	                &		        &	&	4537.94	&	4538.93	&	\\
        &       $	4\nu_4	                $	&	                &		        &	&	5371.39	&	5364.56	&	\\
        &       $	2\nu_2+3\nu_4^3	        $	&	                &		        &	&	5475.84	&	5480.07	&	\\
        &	$	4\nu_2+2\nu_4	        $	&	                &		        &	&	5601.91	&	5607.20	&	\\
$ E'$	&	$	 \nu_4^1	        $	& \onlinecite{nu4_94}	& 1397.0\phantom{00}	&	&	1385.99	&	1387.26	&	\\
        &	$	2\nu_2+\nu_4^1	        $	&	                &		        &	&	2688.80	&	2691.61	&	\\
        &	$	2\nu_4^2	        $	&	                &	 	        &	&	2759.77	&	2762.05	&	\\
        &	$	 \nu_3^1	        $	& \onlinecite{nu3_97b}	& 3160.8\phantom{00}	&	&	3158.88	&	3158.83	&	\\
        &	$	3\nu_4^1	        $	&	                &		        &	&	4074.69	&	4075.46	&	\\
        &	$	2\nu_2+2\nu_4^2	        $	&	                &		        &	&	4087.92	&	4091.72	&	\\
$A_2''$	&	$        \nu_2                  $       & \onlinecite{nu1_92}	& 606.453	        &	&	602.43	&	602.43  &	\\
      	&	$       3\nu_2                  $       & 	                & 	                &	&	2010.09	&	2010.09 &	\\
      	&	$        \nu_2+2\nu_4^0         $       & 	                & 	                &	&       3372.27	&	3371.59 &	\\
      	&	$       5\nu_2                  $       & 	                & 	                &	&	3569.96	&	3569.95 &	\\
        &	$        \nu_1+\nu_2            $       & 	                & 	                &	&	3596.35	&	3596.30 &	\\
        &       $       3\nu_4^3                $       & 	                & 	                &	&	4768.70	&	4767.06 &	\\
        &	$        \nu_2+2\nu_4^0         $       & 	                & 	                &	&	4823.32	&	4822.79 &	\\
        &	$        \nu_1+3\nu_2           $       & 	                & 	                &	&	4981.58	&	4981.52 &	\\
$E''$	&	$	 \nu_2+\nu_4^1	        $	&	                &		        &	&	2000.24	&	2002.22	&	\\
        &	$	 \nu_2+2\nu_4^2	        $	&	                &		        &	&	3388.24	&	3391.11	&	\\
        &	$	3\nu_2+\nu_4^1	        $	&	                &		        &	&	3426.45	&	3430.06	&	\\
        &	$	 \nu_2+\nu_3^1	        $	&	                &		        &	&	3736.40	&	3736.97	&	\\
        &	$	 \nu_2+3\nu_4^1	        $	&	                &		        &	&	4726.62	&	4728.62	&	\\
        &	$	3\nu_2+2\nu_4^2	        $	&	                &		        &	&	4835.22	&	4839.85	&	\\
        &	$	 \nu_1+\nu_2+\nu_4^1	$	&	                &		        &	&       4980.92	&	4983.16	&	\\
\hline
\end{tabular}
\end{center}
\begin{flushleft}
\noindent \vspace*{0.1truecm}
$^a$ Experimental values of band centers used to replace the theoretical values $P_{\rm max}=32$, see text.  \\
$^b$ Computed using the $P_{\rm max}=24$ basis set in conjunction with PES-2\cite{aa15}.\\
$^c$ Computed using the $P_{\rm max}=32$ basis set in conjunction with PES-2\cite{aa15}. \\
\end{flushleft}
\end{table}

Table~\ref{t:rot:energies} shows a comparison of the pure rotational energies ($J\le 5$) of CH$_3$ before and after refinement of $r_{\rm e}$ illustrating the importance of this step.

\begin{table}
\caption{\label{t:rot:energies}  Theoretical rotational term
values ($N\le 5$, in \cm) of CH$_3$, computed with TROVE using different
equilibrium structure parameters.}
\begin{center}
\renewcommand{\arraystretch}{0.95}
\begin{tabular}{cccrrrr}
\hline
\hline
\multicolumn{3}{c}{States}  &   \multicolumn{3}{c}{Term values}          \\
\hline
$N$   &  $K$   &$\tau_{\rm rot}$ &  Obs. &  Obs.-Calc.$^a$& Obs.-Calc.$^b$ \\
\hline
1   &   1    &    0    &  14.3189 & 0.032377 &  0.004027 \\
2   &   0    &    1    &  57.4396 & 0.112005 & -0.002023 \\
2   &   2    &    0    &  38.1186 & 0.092340 &  0.017004 \\
2   &   1    &    0    &  52.6112 & 0.106875 &  0.002511 \\
3   &   3    &    0    &  71.3965 & 0.179934 &  0.038989 \\
3   &   2    &    0    &  95.5353 & 0.203902 &  0.014649 \\
3   &   1    &    0    & 110.0032 & 0.219365 &  0.001200 \\
4   &   0    &    0    & 191.2473 & 0.375024 & -0.004034 \\
4   &   4    &    0    & 114.1491 & 0.295456 &  0.070301 \\
4   &   2    &    0    & 172.0038 & 0.353500 &  0.012772 \\
4   &   3    &    0    & 147.9203 & 0.327970 &  0.035289 \\
\hline
\end{tabular}
\end{center}
\begin{flushleft}
\noindent \vspace*{0.1truecm}
$^a$ {Calculated using $r_{\rm e} = 1.07736927$~\AA\ and $\alpha_{\rm e}= 120.0^\circ$ (PES-1, see text).} \\
\noindent \vspace*{0.1truecm}
$^b$ {Calculated using $r_{\rm e} = 1.0762977119$~\AA\ and $\alpha_{\rm e}=120.0^\circ$ (PES-2, see text).}
\end{flushleft}
\end{table}



The vibrational transition moments are defined as
\begin{equation}\label{e:vib:mu}
 \mu_{V'V} = \sqrt{\sum_{\alpha=x,y,z} | \langle V'\mid \bar\mu_\alpha \mid V \rangle |^2 }
\end{equation}
 where $|V'\rangle$ and $|V\rangle$ denote $J=0$ vibrational wavefunctions and $\bar\mu_\alpha$ is the electronically-averaged dipole moment in the molecular frame (see the section entitled `Dipole moment surface' above).
 For calculation of vibrational transition moments we used our \abinitio\ PES-1 and truncated the vibrational basis set at polyad number $P_{\rm max}=32$.
A number of computed transition moments for the strongest lower lying bands are listed in Table 3 where they are compared with the available experimental data.
 The complete list of theoretical transition moments is given as Supporting Information and can be also found at \url{www.exomol.com}.



\begin{table}
\caption{\label{t:tm:1}  Band Centers $\nu_{fi}$ and Vibrational Transition Moments $\mu_{fi}$ for CH$_3$.$^a$}
\begin{center}
\renewcommand{\arraystretch}{1.25}
\begin{tabular}{llllllll}
\hline
\multicolumn{2}{c}{States}       &  \multicolumn{1}{c}{$\nu_{fi}$/\cm} &
\multicolumn{2}{c}{Calc. $\mu_{fi}$/D} & Obs. $\mu_{fi}$/D
   & Ref.    \\
\hline
$f$                            & $i$   \\
\hline
$ 2\nu_2	        $ &   $\nu_2$	 &	678.81	 &	0.25684	  && 0.31(6)   & [\citenum{mu2_08} ] 	           \\
$  \nu_2	        $ &	0        &	602.43	 &	0.20403	  && 0.215(25) & [\citenum{mu2_05,mu2_83,mu2_89} ] \\
$  \nu_3^1	        $ &	0	 &	3158.83	 &	0.03999	  && 0.03(27)  & [\citenum{mu3_95,mu3_94} ] 	   \\
\cline{1-8}
$  \nu_4^1	        $ &	0	 &	1387.26	 &	0.02931	  &&	       &		                   \\
$ 2\nu_3   +  \nu_4	$ &	0	 &	4529.74	 &	0.02049	  &&	       &		                   \\
$  \nu_1   +  \nu_4^1	$ &	0	 &	4383.56	 &	0.00866	  &&	       &		                   \\
$ 2\nu_2   +  \nu_3^1	$ &	0	 &	4396.18	 &	0.00486	  &&	       &		                   \\
$ 2\nu_3^2	        $ &	0	 &	6294.76	 &	0.00462	  &&	       &		                   \\
$  \nu_1   +  \nu_3^1	$ &	0	 &	6076.68	 &	0.00321	  &&	       &		                   \\
$ 2\nu_4^2	        $ &	0	 &	2762.05	 &	0.00313	  &&	       &		                   \\
$  \nu_3^1 + 2\nu_4	$ &	0	 &	5864.94	 &	0.00242	  &&	       &		                   \\
$ 3\nu_4^1	        $ &	0	 &	4075.46	 &	0.00186	  &&	       &		                   \\
$ 2\nu_2+\nu_3^1+\nu_4^1$ &	0	 &	5789.16	 &	0.00130	  &&	       &		                   \\
$ 4\nu_2	        $ &	0	 &	5856.39	 &	0.00116	  &&	       &		                   \\
\hline
\end{tabular}
\end{center}
\begin{flushleft}
\noindent \vspace*{0.1truecm}
$^a$ The transitions originate in the vibrational ground state
($i=0$) with the exception of the hot band $2\nu_2\leftarrow\nu_2$.
\end{flushleft}
\end{table}

\subsection{Intensity simulations} \label{subsec:intens}
The simulation of absorption spectra at a given temperature $T$ and within a
particular wavenumber interval requires knowledge of  the upper and lower-state energies and the
Einstein coefficients $A(f \leftarrow i)$ [or the line strengths
$S(f \leftarrow i)$; the relationship between
 $A(f \leftarrow i)$ and
$S(f \leftarrow i)$ is described in Ref.~\citenum{dipole05}] for
all transitions in the chosen wavenumber range. In practice, however, the transitions with intensities below a chosen limit are discarded. The
most straightforward presentation of the spectral data is a `stick' diagram with the stick heights representing the integrated absorption coefficients from
Eq.~\eqref{e:intensity-absorption}. We report here such
simulations for the CH$_3$ absorption bands in the wavenumber interval 600--1200~\cm\ for the out-of-plane bending mode $\nu_2$. The line strengths in
Eq.~\eqref{e:intensity-absorption} are computed from
Eq.~\eqref{e:linestrength_deg} with the spin statistical weights
$g_{\rm ns}$ from Ref.~\citenum{aa15}.
The simulations are based on PES-2 and the computed DMS
described above.
The partition-function value used was $Q$ $=$ 732.734,
obtained at 300~K as a summation over all variational term values ($J\le 40$) below
36871.73~\cm.  We have computed
2,058,655,166 transitions using the GPU GAIN-MPI program\citep{17AlYuTe}
within the various limits defined above.

\begin{figure}[h]
\begin{center}
\caption{\label{fig:overview} An overview of the absorption spectrum (cross sections) of CH$_3$ at different temperatures $T=$ 300, 500, 1000 and 1500~K generated using our line list and the Gaussian line profile with the full-width-at-half-maximum of 1~\cm.    }
\includegraphics[width=0.9\textwidth]{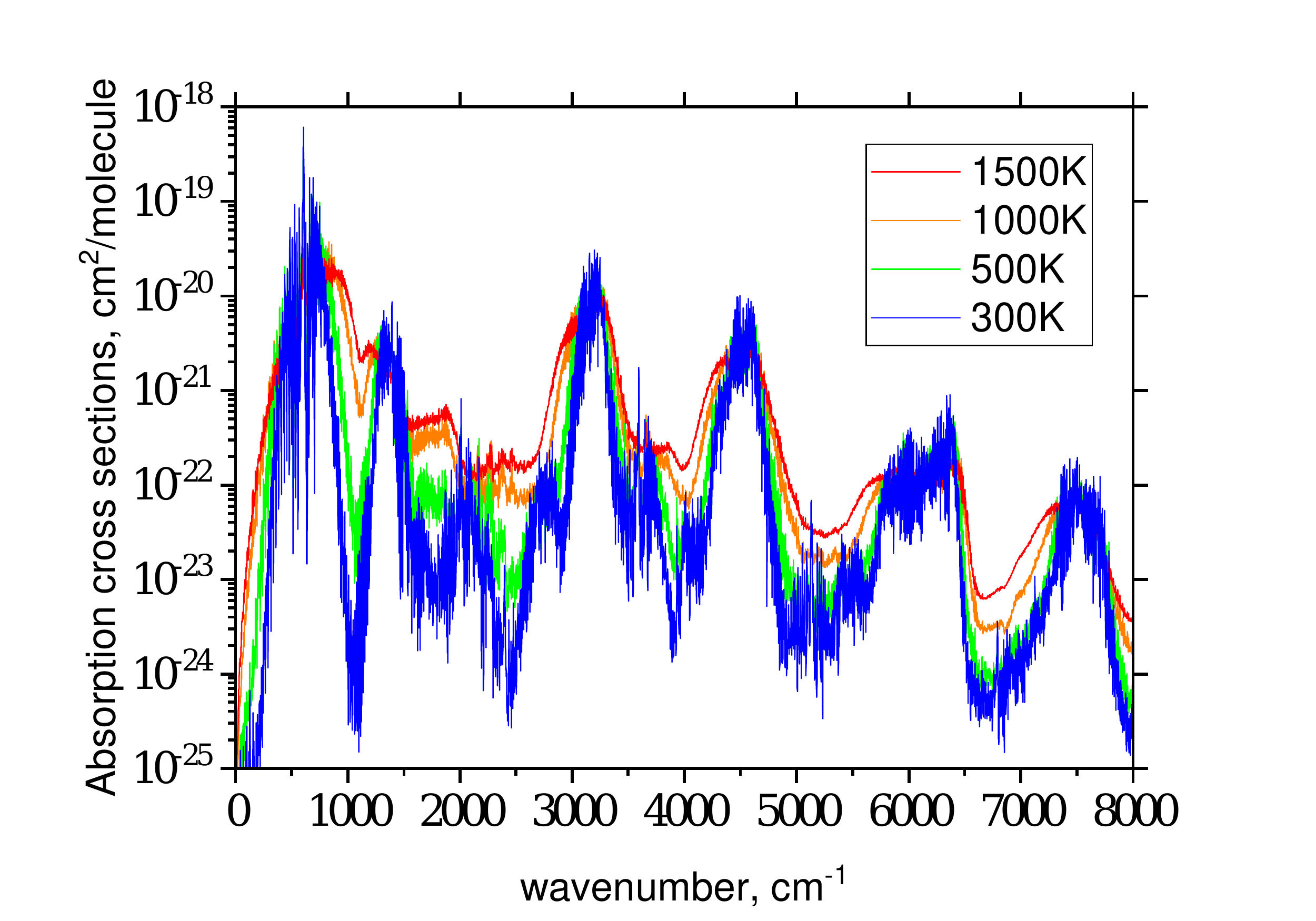}
\end{center}
\end{figure}

\begin{figure}[h]
\begin{center}
\caption{\label{fig:296K} A selection of the strongest absorption bands of CH$_3$ at $T=296$~K generated using the line list.  A Gaussian line profile with the half-width-half-maximum of 0.08~\cm\ was used in production of the cross sections shown.  }
\includegraphics[width=0.9\textwidth]{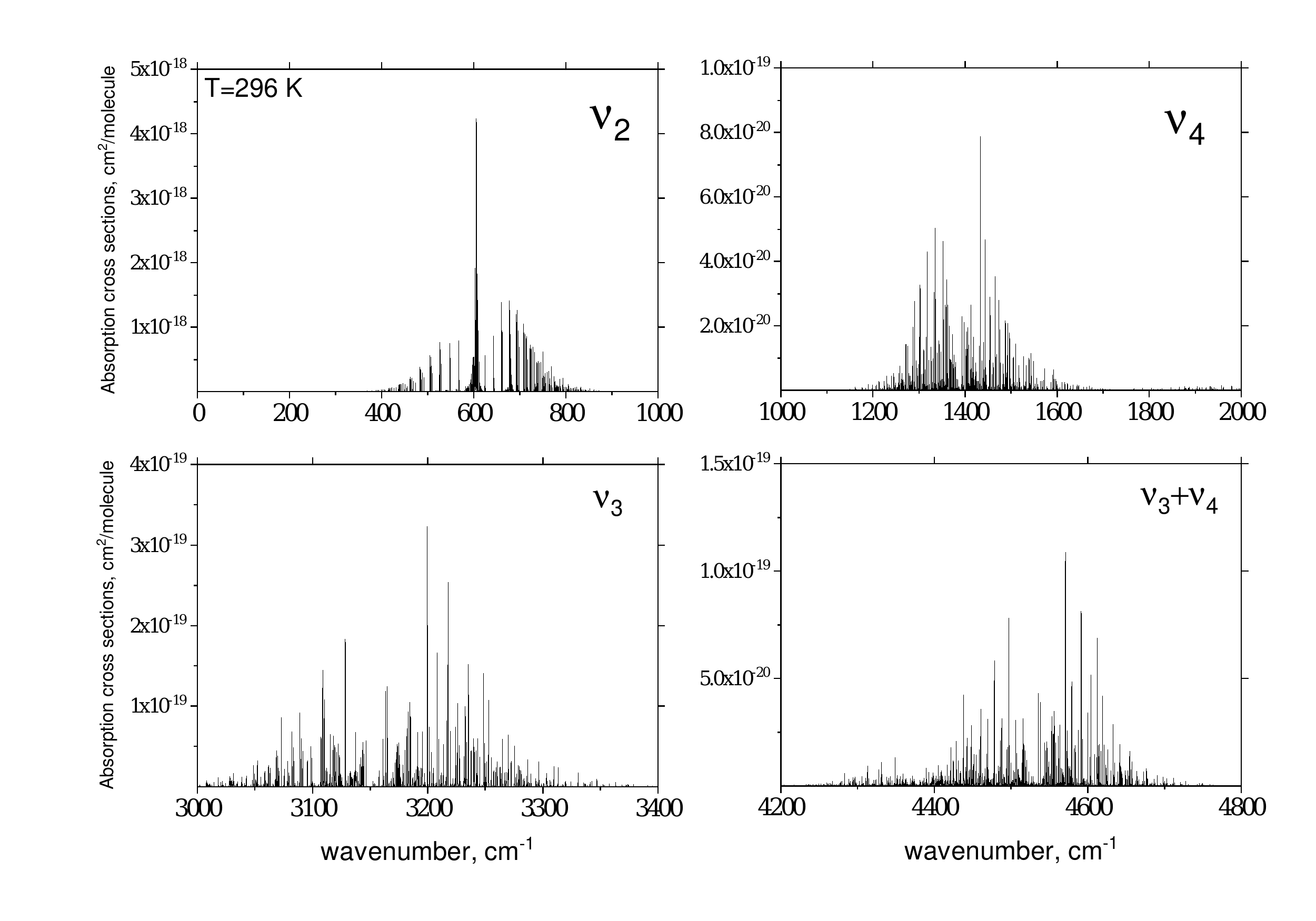}
\end{center}
\end{figure}

Figure~\ref{fig:overview} gives an overview (log-scale) of the absorption spectrum of CH$_3$ at different temperatures produced using the line list (log-scale) by means of the ExoCross program\cite{exocross}. Figure~\ref{fig:296K} shows  four regions with the strongest, dipole-allowed bands $\nu_2$, $\nu_4$, $\nu_3$ and $\nu_3+\nu_4$.

\begin{figure}[h]
\begin{center}
\caption{\protect\label{fig:nu2}
Emission spectra of CH\3.
(a)
Rovibrational simulation assuming the CH$_3$ radicals to be in
thermal equilibrium at
 $T$ $=$ 300~K.
(b)
CH$_3$ ($\nu_2$) out-of-plane bending mode
emission spectrum (dots) obtained\protect\cite{nu2_82a}
after
 dissociation of
CH$_3$I $\rightarrow$ CH$_3$($\nu_2$) + I$^*$.
  The continuous curve is the best fit\protect\cite{nu2_82a}
 involving the hot bands
 $(v_2+1)\, \nu_2$ $\leftarrow$
 $v_2\, \nu_2$ with
 $v_2$ $\leqslant$ 9. The spectrum is
convolved with a
 spectrometer slit function with a FWHM of
33~\cm\ and a 19~\cm\ bandwidth ascribed to the breadth of the $\Delta K$ $=$ 0 manifold of transitions with varying $J$ values.
Reproduced from
Ref.~\citenum{nu2_82a}
 with the permission of AIP Publishing.
(c)
Vibrational simulations
at temperatures $T$ $=$ 1000, 2000, and 3000~K, respectively, taking into account the vibrational
transitions
 $(v_2+1)\, \nu_2$ $\leftarrow$
 $v_2\, \nu_2$ with
 $v_2$ $\leqslant$ 9 (see text).
}
\includegraphics[width=0.8\textwidth]{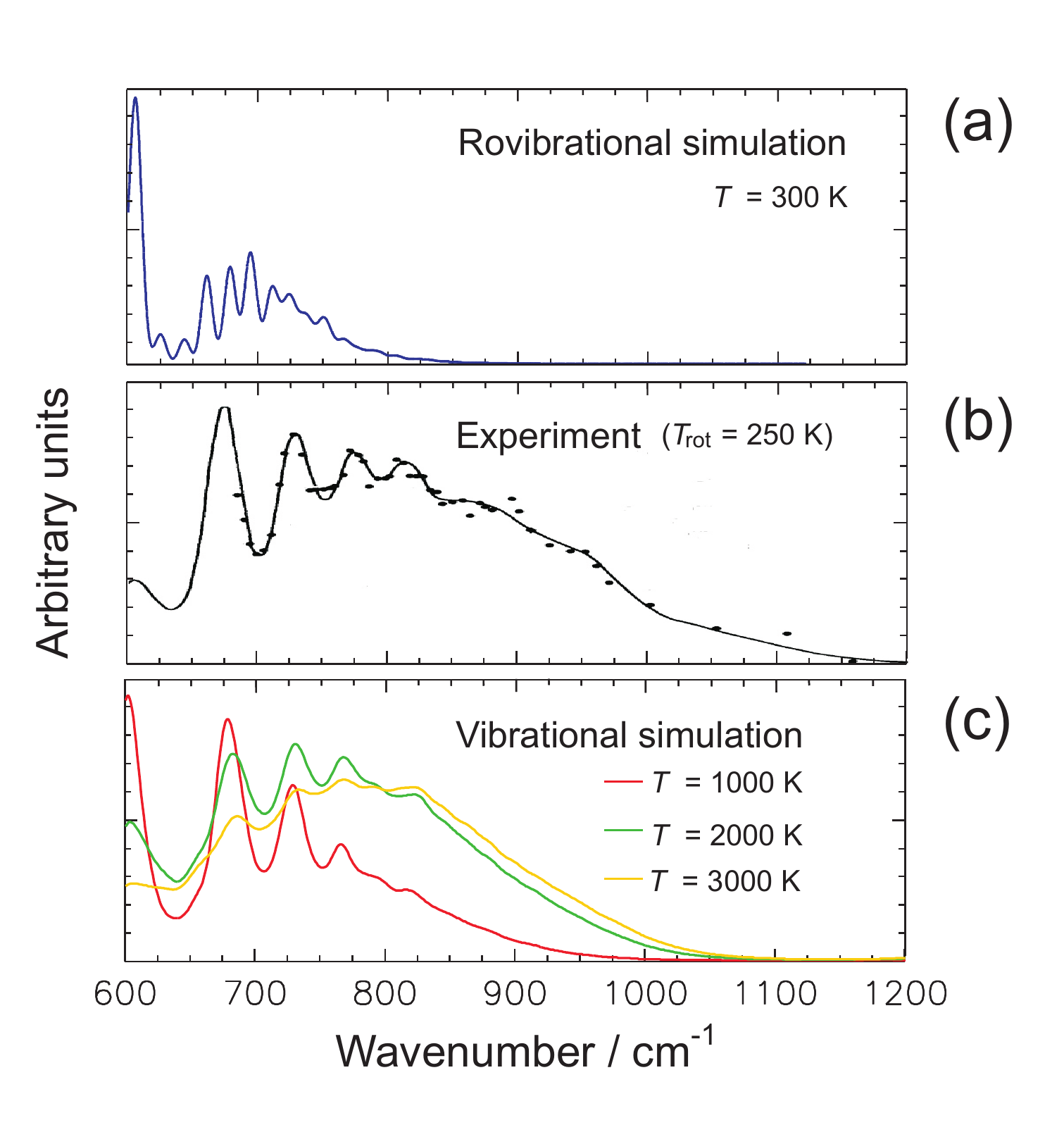}
\end{center}
\end{figure}

Figure~\ref{fig:nu2} shows the emission spectrum of CH$_3$ in
the wavenumber interval 600-1200~\cm,
simulated with TROVE
at two different levels of theory,
rovibrational and purely vibrational. The simulated spectra are compared to
an experimental spectrum recorded by
Hermann and Leone\cite{nu2_82a,nu2_82b} (see Fig.~4 of Ref.~\citenum{nu2_82a}).

Figure~\ref{fig:nu2}(a)  shows a simulation of the CH$_3$ emission spectrum, computed with
TROVE at a temperature of $T$ $=$ 300~K, taking into account all vibrational transitions
in the 600-1200~\cm\ interval that are calculated with the $P_{\rm max} = 32$ basis set. The simulation involves the convolution of the calculated intensities with a Gaussian line shape function with a full width at half maximum (FWHM) of 17~cm$^{-1}$.

The simulation is compared with the experimental spectrum obtained by
Hermann and Leone\cite{nu2_82a,nu2_82b}, shown in
Figure~\ref{fig:nu2}(b).\footnote{Reproduced from
Hermann, H. W.; Leone, S. R. The Journal of Chemical Physics 1982, 76, 4759--4765,
 with the permission of AIP Publishing.}
 In the experiment, the CH$_3$ radicals
were
produced in a photo-fragmentation process
of methyl iodide CH$_3$I.
Hermann and Leone\cite{nu2_82a,nu2_82b} suggested that after the breaking of the C--I bond,
 the CH$_3$ radicals are predominantly produced in excited states of the out-of-plane bending mode
 $\nu_2$. The CH$_3$ fragment of a dissociated CH$_3$I molecule will typically have a pyramidal structure close
to that of the methyl group in
  CH$_3$I. For a CH$_3$ radical, however, which is planar at equilibrium, such structures are associated with
high excitations of the $\nu_2$ vibrational mode.
 Excitations with $v_2$ $\leqslant$  10 have been
observed in the
emission experiment
of Refs.~\citenum{nu2_82a,nu2_82b}.

In order to simulate vibrationally very hot transitions (hotter than 1500~K) corresponding to
the experimental spectrum of
Refs.~\citenum{nu2_82a} which involves out-of-plane bending states
 $v_2\, \nu_2$ with $v_2$ $\leqslant$  10, we have carried out simulations employing
a so-called one-band model (with the one band being the $\nu_2$ fundamental band here). In the one-band
model, we use the $\nu_2$-band data from the
`parent',
 300~K
 ro-vibrational line list also for the hot bands accompanying the
 $\nu_2$ band. The procedure employed is detailed in
 Ref.~\citenum{jt729} and we outline it briefly here:
We initially produce the
 300~K, $\nu_2$-band cross sections by means of the ExoCross program\cite{exocross} and convolve them
with a Gaussian profile of   FWHM $=$ 17~cm$^{-1}$ between 0 and 1200~cm$^{-1}$, generating 1201 data points. Then the
wavenumbers of the computed cross sections are shifted by
$-$606.4531~cm$^{-1}$, positioning the band center at zero, and the cross sections are normalized.
 A local version of ExoCross now obtains,
 from             the vibrational transition moments computed with TROVE,
the vibrational band intensities
for the hot bands
 $(v_2+1)\, \nu_2$ $\leftarrow$
 $v_2\, \nu_2$,
 $v_2$ $\leqslant$ 9. Finally, the simulated spectrum is generated  by placing,
for each hot band,
 the $\nu_2$-band profile
at the band center $\nu_{fi}$ of the hot band in question, scaled by its
vibrational band intensity.
Such simulations have been carried out for
 temperatures $T$ of  1000, 2000, and 3000~K, respectively, and the results are shown in
Figure~\ref{fig:nu2}(c),
   where they can be compared to the experimental results\cite{nu2_82a,nu2_82b}
in
Figure~\ref{fig:nu2}(b).

Figure~\ref{fig:nu2}   shows that the `standard'
 $T$~$=$~300~K rovibrational simulation of the CH$_3$ emission spectrum
[Figure~\ref{fig:nu2}(a)]
has little resemblance to
the experimental spectrum from
  Ref.~\citenum{nu2_82a}
[Figure~\ref{fig:nu2}(b)]. Obviously in the experiment, the CH$_3$ molecules populate states
of much higher energies than those accessed in thermal equilibrium at
 $T$~$=$~300~K. However, among the `vibrational simulations' in
Figure~\ref{fig:nu2}(c), the curve obtained for
 $T$~$=$~2000~K has a very substantial similarity to the experimental curve. This confirms the
suggestion by
Hermann and Leone\cite{nu2_82a,nu2_82b} that dissociation of
 CH$_3$I  produces
 CH$_3$ radicals in highly excited states of the out-of-plane bending mode $\nu_2$.
The successful simulation of the emission spectrum of
Refs.~\citenum{nu2_82a,nu2_82b} lends credibility to the
 \abinitio\ DMS of the present work; the
intensities based on
this DMS are in very good qualitative agreement with experiment.

Our complete $T=1500$~K CH$_3$ line list is accessed via the repository  \url{www.zenodo.org}, see Ref.~\citenum{19AdYaYu.zenodo}.
It provides transition energies, line strengths, Einstein
coefficients $A(f \leftarrow i)$ and the temperature dependent partition function $Q(T)$. We expect the line list to be applicable for  temperatures below 1500~K. However, the simulated spectra will become increasingly inaccurate with increasing temperature. The line list is given in the ExoMol format\cite{exomol} which can be used together with the ExoCross program\cite{exocross} to generate spectra of CH$_3$.

\section{Conclusion}

We report here simulations of spectra for the methyl radical, extending over a significant portion  of the infrared spectral region.
The positions and intensities calculated for  the transitions
are in
excellent agreement with experiment, as demonstrated by
detailed comparisons with observed room-temperature spectra.

The CH$_{3}$ line list of the present work
 will facilitate
 detections of the methyl radical in space.
In the present work we have generated, refined, and validated the
potential energy and dipole moment surfaces required for the spectral simulations, and we have established
the level of accuracy attainable in variational nuclear-motion
calculations with our computational resources.
We have produced a methyl radical line list consisting
of 2 billion transitions between 9,127,123 energy levels for ro-vibrational
states up to $J_{\rm max}$ = 40 and energies up to 19~000~cm$^{-1}$.

\clearpage

\begin{acknowledgement}
A.Y. acknowledges support from DESY (HGF IVF) and from the Cluster of Excellence 'The Hamburg Centre for Ultrafast Imaging' of the Deutsche Forschungsgemeinschaft (DFG) - EXC 1074 - project ID 194651731.
S.Y. is grateful for support from the UK Science and Technology Research Council (STFC)
ST/R000476/1. This work made extensive use of UCL's Legion high performance (HPC) computing facilities as well as of HPC provided by DiRAC supported by STFC and BIS.
It was further supported in part by  grant JE 144/25-1 from the DFG.
\end{acknowledgement}

\begin{suppinfo}
The Supporting Information for this work includes: (i) dipole moment parameters $\mu_{k,l,m,\ldots}^{(\Gamma)}$,  (ii) potential energy parameters
$f_{jk\dots}^{(s)}$; (iii) Fortran routines for calculating the dipole moment and potential energy values for a given geometry.  Our complete $T=1500$K CH$_3$ line list together with the partition function can be accessed via the Zenodo repository, \url{www.zenodo.org}~\cite{19AdYaYu.zenodo} as well as at \url{www.exomol.com}.

\end{suppinfo}


\clearpage

\clearpage
\begin{center}
FIGURE CAPTIONS
\end{center}

\begin{list}{}{\leftmargin 2cm \labelwidth 1.5cm \labelsep 0.5cm}
\item[Figure~\ref{fig:rms-DMS}]
Root-mean-square (RMS) errors of the fittings to the \ai\ dipole moment values.
The results of a series of fittings are shown. In each fitting, the data set includes the \ai\ points with
electronic energy (relative to the potential energy minimum) $V$ $\leqslant$ $E$ (see text).
\item[Figure~\ref{fig:overview}]  An overview of the absorption spectrum (cross sections) of CH$_3$ at different temperatures $T=$ 300, 500, 1000 and 1500~K generated using our line list and the Gaussian line profile with the full-width-at-half-maximum of 1~\cm.
\item[Figure~\ref{fig:296K}] A selection of the strongest absorption bands of CH$_3$ at $T=296$~K generated using the line list.  A Gaussian line profile with the half-width-half-maximum of 0.08~\cm\ was used in production of the cross sections shown.
\item[Figure~\ref{fig:nu2}]
Emission spectra of CH\3.
(a)
Rovibrational simulation assuming the CH$_3$ radicals to be in
thermal equilibrium at
 $T$ $=$ 300~K.
(b)
CH$_3$ ($\nu_2$) out-of-plane bending mode
emission spectrum (dots) obtained\protect\cite{nu2_82a}
after
 dissociation of
CH$_3$I $\rightarrow$ CH$_3$($\nu_2$) + I$^*$.
  The continuous curve is the best fit\protect\cite{nu2_82a}
 involving the hot bands
 $(v_2+1)\, \nu_2$ $\leftarrow$
 $v_2\, \nu_2$ with
 $v_2$ $\leqslant$ 9. The spectrum is
convolved with a
 spectrometer slit function with a FWHM of
33~\cm\ and a 19~\cm\ bandwidth ascribed to the breadth of the $\Delta K$ $=$ 0 manifold of transitions with varying $J$ values.
Reproduced from Fig.~4 of
Ref.~\citenum{nu2_82a}
 with the permission of AIP Publishing.
(c)
Vibrational simulations
at temperatures $T$ $=$ 1000, 2000, and 3000~K, respectively, taking into account the vibrational
transitions
 $(v_2+1)\, \nu_2$ $\leftarrow$
 $v_2\, \nu_2$ with
 $v_2$ $\leqslant$ 9 (see text).
\end{list}



\clearpage



\providecommand{\latin}[1]{#1}
\makeatletter
\providecommand{\doi}
  {\begingroup\let\do\@makeother\dospecials
  \catcode`\{=1 \catcode`\}=2 \doi@aux}
\providecommand{\doi@aux}[1]{\endgroup\texttt{#1}}
\makeatother
\providecommand*\mcitethebibliography{\thebibliography}
\csname @ifundefined\endcsname{endmcitethebibliography}
  {\let\endmcitethebibliography\endthebibliography}{}

\clearpage

\begin{center}
{\Large TOC graphics} \\[36pt]
\includegraphics[width=0.9\textwidth]{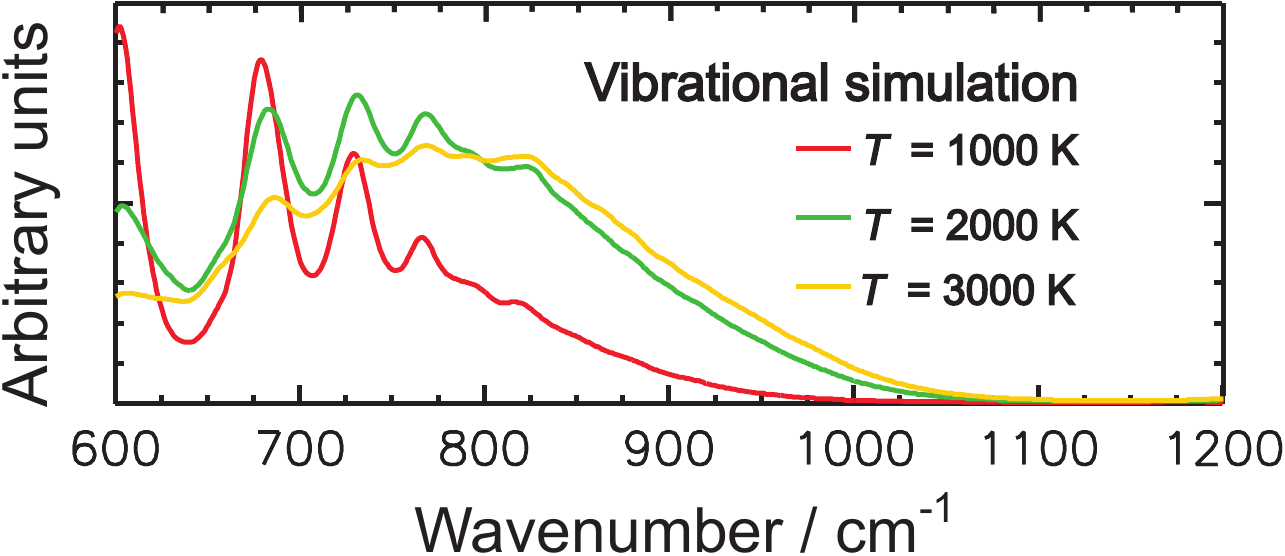}
\end{center}

\end{document}